\documentclass[conference,a4paper]{IEEEtran}
\usepackage[dvips]{graphicx}
\usepackage{amsmath,amssymb}

\newcommand{\defeq}{\stackrel{\triangle}{=}}

\newtheorem{lemma}{Lemma}
\newtheorem{theorem}{Theorem}

\newcommand{\qed}{\fbox{}}
\newif\ifhoge
\hogetrue

\begin{document}

\sloppy

\title{Bitwise MAP Algorithm for Group Testing based on Holographic Transformation}

\author{
  \IEEEauthorblockN{Tadashi Wadayama and Taisuke Izumi}
  \IEEEauthorblockA{Department Computer Science and Engineering, \\Nagoya Institute of Technology,
    Nagoya,  Japan\\
    Email: wadayama@nitech.ac.jp, t-izumi@nitech.ac.jp} 
}



\maketitle

\begin{abstract}

In this paper, an exact bitwise MAP (Maximum A Posteriori) estimation algorithm for group testing problems is presented.
We assume a simplest non-adaptive group testing scenario including 
$N$-objects with binary status and $M$-disjunctive tests.
If a group contains a positive object, the test result for the group is assumed to be one;
otherwise, the test result becomes zero.
Our inference problem is to evaluate the posterior probabilities of the objects from the observation of
$M$-test results and from our knowledge on the prior probabilities for objects.
If the size of each group  is bounded by a constant,  a naive inference algorithm requires $O(N 2^N)$-time
for computing the posterior probabilities for objects. 
Our algorithm runs with 
$O(N^2 2^M)$-time, which is exponentially faster than the naive inference algorithm 
under a common situation with $M << N$. 
The heart of the algorithm is the {\em dual expression} of the posterior values.
The derivation of the dual expression 
can be naturally described based on a holographic transformation to the normal factor graph (NFG)
representing the inference problem. In order to handle OR constraints in the NFG, we introduce 
a novel holographic transformation that converts an OR function to a function similar to 
an EQUAL function.
\end{abstract}

\section{Introduction}

Graphical models, such as factor graphs and
normal factor graphs \cite{Forney1} \cite{Loeliger1}, 
can provide a concise description of the probabilistic assumption of an inference problem
and they are indispensable for analyzing message passing inference algorithms 
such as BP (Belief Propagation). 
For example, the relationship between ``codes and graphs'' is one of key concepts 
of modern coding theory.

Al-Bashabsheh and Mao \cite{Al-Bashabsheh} recently shed a new light to 
normal factor graphs. They clearly showed that 
{\em holographic transformations} to normal factor graphs 
are versatile tools for deriving non-trivial identities on the partition function of a normal factor graph \cite{Forney2}.
A holographic transformation is a local graphical transformation 
that preserves the partition function. 
It should be remarked that the holographic transformation has been used in many research fields.
The prominent example is the class of holographic algorithms invented by Valiant \cite{Valiant}.
He showed that several combinatorial enumeration problems defined on planer graphs can be transformed 
into perfect matching problems via appropriate holographic transformations.
Such a planar perfect matching problem is solvable in polynomial time.
Another example is duality theorems \cite{Forney1} \cite{Forney2} \cite{Mao} for codes defined on graphs.

The main contribution of this paper is 
a non-trivial expression, that is called {\em dual expression}, 
of the posterior values for  a non-adaptive group testing problem.
The derivation is based on a holographic transformation to 
the normal factor graph representing a group testing inference problem.
The derivation process has  similarities to the proof of MacWilliams identity \cite{Forney2}
and a bitwise MAP decoding algorithm by Hartmann and Rudolph \cite{Hartmann}  
for binary linear codes. However, in our case, we cannot rely on the standard Fourier ({\it i.e.,} Hadamard)
transformation because we need to treat OR constraints instead of even parity constraints.
A local linear transformation matched to OR constraints plays a key role for the following discussion.

\section{Preliminaries}

\subsection{Inference on group testing problems}

The research of group testing started from the celebrated work by Dorfman \cite{Dorfman} and 
has been extensively studied \cite{Du1}.
We here suppose the following simplest setting for a non-adaptive group testing.
Assume that we have $N$-{\em objects} and some groups of these objects.
Each object can take value 1 (positive)  or 0 (negative) according to the prior probability for 
each object. A {\em test} can be applied for each predetermined group. The result of 
a test is positive if  the group contains a positive object; otherwise the test result becomes
negative.  Our inference problem is to evaluate the posterior probabilities for objects from 
$M$-disjunctive test results and from our knowledge on the prior probabilities.
Development of fast inference algorithms evaluating posterior probabilities (or their estimates)  
is an active area of research;
for example, see  approximate inference algorithms based on BP \cite{Kanamori} \cite{Sejdinovic}.

\subsection{Problem setup}
Let $S_j (j \in [1,N])$ be a binary (zero or one) 
independent random variable representing the state of the $i$-th object ({\em i.e.,} negative or positive).
The notation $[a,b]$ represents consecutive integers from $a$ to $b$.
The vector of the random variables $\mathbf{S} \defeq (S_1,\ldots, S_N)$ is thus distributed 
according to the joint distribution: 
\begin{equation}
P_{\mathbf{S}}(s_1,\ldots, s_N) = \prod_{j=1}^N P_{S_j}(s_j),
\end{equation}
where $(s_1,\ldots, s_N) \in \{0,1\}^N$. We suppose that 
an inference algorithm perfectly knows 
these prior probabilities $P_{S_j}(s_j)$.
Assume that an undirected bipartite graph $G \defeq (V_1,V_2, E)$,  called a {\em pooling graph}, is given where
$V_1 \defeq \{v^{(1)}_1,v^{(1)}_2,\ldots, v^{(1)}_N\}$ and $V_2 \defeq \{v^{(2)}_1,v^{(2)}_2,\ldots, v^{(2)}_M\}$ 
are sets of vertices, and $E$ is the set of edges connecting a vertex in $V_1$ and a vertex in $V_2$,
namely $E \subset \{(v^{(2)}_i,v^{(1)}_j) \in V_2 \times V_1\}$.
The set $\sigma(i)$ is defined by
\begin{equation}
\sigma(i)   \defeq \{j \in [1,N] \mid  (v^{(2)}_i, v^{(1)}_{j}) \in E  \}  
\end{equation}
for $i \in [1,M]$.
The Boolean function $OR: \{0,1 \}^r \rightarrow \{0,1\}$ is just the logical OR function 
with $r$-inputs defined as 
\[
OR(x_1,\ldots, x_r) = \Bbb I[\exists k \in [1,r], x_k=1 ].
\]
The indicator function $\Bbb I[condition]$ takes the value one if the condition is true; 
otherwise it takes the value zero.

A binary random variable $T_i (i \in [1,M])$ representing a test result
is defined by 
$
T_i = OR(S_k|_{k \in \sigma(i)}), i \in [1,M],
$
where the notation $x_{k} \mid_{k \in \{i_1,\ldots, i_k\}}$ represents a sequence of variables
$x_{i_1}, \ldots, x_{i_k}$. 
The vector composed from $T_i (i \in [1,M])$ is denoted by $\mathbf{T} \defeq (T_1,\ldots, T_M)$.
It is evident that there is one-to-one correspondence between $S_j$ and $v^{(1)}_j \in V_1$
and also between $T_i$ and $v^{(2)}_i \in V_2$. The index set $\sigma(i)$ represents a group 
corresponding to the $i$-th test result.

Assume that we observed $\mathbf{t} \defeq (t_1,t_2,\ldots, t_M) \in \{0,1\}^M$ as a realization of $\mathbf{T}$.
Our goal is to evaluate the log posterior probability ratio defined by
\[
R_\ell \defeq 
\log
({P_{S_\ell |\mathbf{T}}(1| \mathbf{t})}/{P_{S_\ell |\mathbf{T}}(0| \mathbf{t})}), \ell \in [1,N].
\]
The probability $P_{S_\ell |\mathbf{T}}(b| \mathbf{t}) (b \in \{0,1\})$ is the posterior probability 
on $\ell$-th object.
From $R_\ell$, we can obtain an estimate vector $\mathbf{\hat s} \defeq (\hat s_1,\ldots, \hat s_N)$ defined by
$
\hat s_\ell \defeq \Bbb I[R_\ell  \ge 0] (\ell \in [1,N])
$
where this estimation rule can be seen as the bitwise MAP estimation rule.
It may be reasonable to consider the bitwise MAP estimation for this group testing problem because 
bitwise MAP estimation  minimizes the bitwise estimation error probability.
By using Bayes' theorem, the posterior probability can be rewritten as
\begin{equation} \label{exponential}
P_{S_\ell |\mathbf{T}}(b| \mathbf{t})  
= 
\frac{1}{Z} \sum_{s_1,\ldots, s_N}
P_{\mathbf{T} |\mathbf{S}}(\mathbf{t}| \mathbf{s}) P_{\mathbf{S}}(\mathbf{s}) \Bbb I[s_\ell = b],
\end{equation}
where $Z$ is just a normalization constant and $\mathbf{s}=(s_1,\ldots, s_N)$. 
As a simplified notation,  
if the domain of the variable is missing in a summation,
all the possible values in the domain is taken to evaluate the sum. 
As in many similar bitwise MAP estimation problems,  naive evaluation according to 
(\ref{exponential}) requires exponential time with the number of variables $N$ to marginalize 
all the variables $s_1,\ldots, s_N$; namely computation time is $O(N 2^N)$ if 
the maximum size of $\sigma(i)$ is bounded by a constant.
This  prohibitive time complexity is the high burden to exploit  the bitwise MAP estimation on this problem.

\subsection{Shrinking pooling graph}
Although it is still exponential time, the exponent of computation time can be greatly reduced 
if we are aware of the following simple fact.
\begin{lemma}[Node elimination]
If $t_i = 0$, then we have $R_k = -\infty$
for any $k \in \sigma(i)$.
\end{lemma}
Proof: If  $t_i = 0$, then $S_k$ should be 0 for any $k \in \sigma(i)$ because of 
the relation $T_i = OR(S_k|_{k \in \sigma(i)})$. This means that
$P_{S_k | \mathbf{T}}(0| \mathbf{t})$ is exactly one.
\hfill\qed

In other words, the lemma states  that all the objects in the group $\sigma(i)$ have the value zero
only if $i$-th test result $t_i$ is zero.
This trivial but useful lemma can significantly reduce the problem size
if the number of  negative objects are small.
Therefore, it might be better to redefine the reduced size problem for a given observation
vector $\mathbf{t}=(t_1,\ldots, t_N)$ as follows.
Let $G^* \defeq (V_1^*, V_2^*, E^*)$ be the induced subgraph of $G$ 
where the vertices of $V_1^*$ and  $V_2^*$ are given by
\begin{eqnarray}
V_1^* &\defeq&  V_1 \backslash \left( \bigcup_{i \in [1,M]: t_i = 0}\{ v_k^{(1)} \in V_1  \mid k \in \sigma(i) \} \right), \\
V_2^* &\defeq& \{v^{(2)}_i \in V_2 \mid t_i = 1 \}.
\end{eqnarray}
In other words, we can exclude the groups whose test result is zero in $V_2$
and its incident nodes in $V_1$ for  evaluating the posterior probabilities.
For the following analysis, it would be convenient to rename the vertices 
in $V_1^*$ and $V_2^*$  as 
\begin{eqnarray}
V_1^* &=& \{v_{k_1}^{(1)}, \ldots,  v_{k_n}^{(1)}\} = \{z_1^{(1)}, \ldots, z_n^{(1)}\}, \\
V_2^* &=& \{v_{l_1}^{(1)}, \ldots,  v_{l_m}^{(1)}\} = \{z_1^{(2)}, \ldots, z_m^{(2)}\}
\end{eqnarray}
and $E^* \subset \{(z^{(2)}_i,z^{(1)}_j) \in V_2^* \times V_1^* \}$.
The random variable corresponding to $z_j^{(1)}$  and $z_i^{(2)}$ 
are denoted by $X_j (j \in [1,n])$ and $Y_i (i \in [1,m])$, respectively.

\begin{figure}[htbp]
\begin{center}
\includegraphics[scale=0.3]{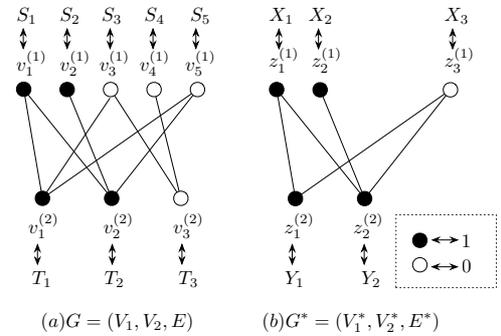}
\end{center}
\caption{Definition of $G$ and $G^*$}
\label{ex1}
\end{figure}

Figure \ref{ex1} illustrates an example of a pair $G$ and $G^*$.
The original pooling graph is depicted in Fig. \ref{ex1}(a).
In this case, we have the test result $\mathbf{t}=(1,1,0)$ which defines 
the induced subgraph $G^*$ illustrated in Fig. \ref{ex1}(b). 

As in the cases of $\mathbf{S}, \mathbf{T}$,  we introduce similar notation such as 
$\mathbf{Y} = (Y_1, \ldots, Y_m)$, $\mathbf{X} = (X_1, \ldots, X_n)$.
In this problem setting, our goal can be recast as the evaluation of the log posterior probability ratio 
$
r_\ell \defeq 
\log ({P_{X_\ell |\mathbf{Y}}(1| \mathbf{1} )}/{P_{X_\ell |\mathbf{Y}}(0| \mathbf{1} )}), \ell \in [1,n]
$
for given $G^*$. The symbol $\mathbf{1}$ represents the vector that all its components are ones.

\subsection{Sum-product form of posterior probabilities}

In this subsection, we will rewrite the posterior probabilities  in sum-product form 
which is the foundation of the following discussion.

As in the derivation of (\ref{exponential}), 
the posterior probabilities $P_{X_\ell |\mathbf{Y}}(b | \mathbf{1} )$  can be expressed as 
\begin{eqnarray}  \nonumber
P_{X_\ell |\mathbf{Y}}(b | \mathbf{1} )  
&=& 
\frac{1}{Z'} \sum_{x_1,\ldots, x_n}
P_{\mathbf{Y} |\mathbf{X}}(\mathbf{1}| \mathbf{x}) P_{\mathbf{X}}(\mathbf{x}) \Bbb I[x_\ell = b]  \\   \nonumber
&=&
\frac{1}{Z'}
 \sum_{x_1,\ldots, x_n}
\left(\prod_{i=1}^m OR(x_k|_{k \in \alpha(i)}) \right) \\
&\times&  \left(\prod_{j=1}^n P_{X_j}(x_j) \right) \Bbb I[x_\ell = b]  
\end{eqnarray}
for $b \in \{0,1\}$, $\ell \in [1,n]$ and $\mathbf{x} = (x_1,\ldots, x_n)$. 
The symbol $Z'$ represents a normalization constant 
which is independent of the value of $b$.
Note that the two sets, $\alpha(i) (i \in [1,m])$ and $\beta(j) (j \in [1,n])$, are  defined by
\begin{eqnarray}
\alpha(i)  &\defeq& \{j \in [1,n]  \mid  (z_i^{(2)},  z_j^{(1)}  ) \in E^*  \}, \\
\beta(j)    &\defeq& \{i \in [1,m] \mid  (z_i^{(2)},  z_j^{(1)}) \in   E^* \},  
\end{eqnarray}
respectively.

For the following argument,  it is useful to define the quantity $a^{(b)}(\ell) (b \in \{0,1\}, \ell \in [1,n])$  by
\begin{eqnarray} \nonumber
a^{(b)}(\ell)  &\defeq&  \sum_{x_1,\ldots, x_n}
\left(\prod_{i=1}^m OR(x_k|_{k \in \alpha(i)}) \right)  \\  \label{posterior}
&\times& \left(\prod_{j=1}^n P_{X_j}(x_j) \right) \Bbb I[x_\ell = b],
\end{eqnarray}
that is called a {\em posterior value}.
By using these posterior values, 
the log posterior probability ratio $r_\ell$ can be evaluated by taking 
the ratio between the posterior values for zero and one: 
\[
r_\ell 
=
\log ({a^{(1)}(\ell)}/{a^{(0)}(\ell)}).
\]

We will further decompose $a^{(b)}(\ell)$ in (\ref{posterior}) into a finer sum-product form
which will be more suitable for a normal factor graph representation to be described in the next section.
As building blocks of the finer representation, we here introduce $EQ$, and $\phi_i^{(\ell, b)}$ functions 
as follows.  
The Boolean equality function $EQ: \{0,1 \}^r \rightarrow \{0,1\}$ with $r$-inputs
is defined by
\[
EQ(x_1,\ldots, x_r) \defeq \Bbb I[x_1 = x_2 = \cdots = x_r].
\]
The weight function $\phi_j^{(\ell,b)}: \{0,1\} \rightarrow \Bbb R (j \in [1,n], \ell \in [1,n], b \in \{0,1\})$ is given by
\begin{equation}
\phi_j^{(\ell, b)}(x) \defeq
\left\{
\begin{array}{ll}
P_{X_j}(x), & j\ne \ell  \\
P_{X_j}(x) \Bbb I[x = b], &  j =  \ell.
\end{array}
\right.
\end{equation}
By using these set of functions,  $a^{(b)}(\ell)$ can be represented as
\begin{eqnarray}  \nonumber
a^{(b)}(\ell) &=&  \sum_{u_1,\ldots, u_n } \sum_{\Gamma }  \left(\prod_{i=1}^m OR(x_{i,k}|_{k \in \alpha(i)} )  \right)  \\ \label{finerrep}
&\times& \left(\prod_{j =1}^n EQ(x_{k,j}|_{k \in \beta(j)} , u_j) \phi_j^{(\ell,b)}(u_j) \right),
\end{eqnarray}
where $\Gamma$ is the set of new binary variables defined as
\[
\Gamma \defeq \{x_{i,j} \mid i \in [1,m], j \in [1,n], (z_i^{(2)},z_j^{(1)}) \in E^* \}
\]
and $u_1,\ldots, u_n$ are also binary variables.

\section{Normal Factor Graph and Holographic Transformation}

\subsection{Dual expression of posterior value}

The main contribution of this paper 
is the next theorem which gives another expression of the posterior value $a^{(b)}(\ell)$.
It will be the foundation of a novel MAP algorithm described later.
\begin{theorem}[Dual Expression]
\label{dualexpression}
The posterior value $a^{(b)}(\ell)$ can be expressed as
\begin{eqnarray} \nonumber
&&\hspace{-1cm}a^{(b)}(\ell) \\ \nonumber
&=& \hspace{-0.5cm} \sum_{w_1,\ldots, w_m} \left(\prod_{i=1}^m (-1)^{w_i (\#\alpha(i) + 1)} \right) 
\left(\prod_{j=1}^n  \Delta_{j}^{(\ell, b)}(w_k |_{k \in \beta(j)}) \right) , 
\end{eqnarray}
where $\ell \in [1,n], b \in \{0,1\}$ and 
variables $w_1,\ldots, w_m$ are binary variables. The notation $\#\alpha(i)$ represents 
the cardinality of $\alpha(i)$. The function $\Delta_j^{(\ell, b)}: \{0,1\}^r \rightarrow \Bbb R (\ell \in [1,n], b \in \{0,1\})$  
is defined by
\begin{equation}
\Delta_j^{(\ell, b)}(y_1,\ldots, y_r)
\defeq
\left\{
\begin{array}{l}
1, \hfill  y_1 = \cdots = y_r = 0 \\
(-1)^{\sum_{k=1}^r y_k} P_{X_j}(0), \hfill  \mbox{otherwise}.
\end{array}
\right. 
\end{equation}
if $j \ne \ell$.
If $j = \ell$, then $\Delta_j^{(\ell, b)}(y_1,\ldots, y_r)$ is defined as
\begin{equation}
\Delta_j^{(\ell, b)}(y_1,\ldots, y_r) 
\defeq
\left\{
\begin{array}{l}
P_{X_j}(b), \quad  y_1 = \cdots = y_r = 0 \\
(-1)^{\sum_{k=1}^r y_k} P_{X_j}(0) \Bbb I[b=0],\hfill  \mbox{otherwise}.
\end{array}
\right.
\end{equation}
\end{theorem}

In the expression of the posterior value (\ref{finerrep}), the indicator variables $u_1,\ldots, u_n$ corresponding to 
$EQ$ nodes take all 
the possible binary $n$-tuples in the summation. 
On the other hand, in the expression of posterior values in Theorem \ref{dualexpression},  
the indicator variables $w_1,\ldots, w_m$ appeared in 
the summation correspond to $OR$ nodes. We therefore call this expression  
the dual expression.

The proof of this theorem heavily relies on a holographic transformation to the normal factor graph 
of the posterior value $a^{(b)}(\ell)$.
In the next section,  we will discuss an appropriate holographic transformation to 
derive the dual expression.

\subsection{Normal factor graph}

The normal factor graph (NFG) is a graphical representation of a function composed from a 
product of many functions. The precise definition of the NFG can be found in \cite{Loeliger1} \cite{Al-Bashabsheh} 
but we here 
introduce a simplified definition enough for this paper.  The NFG of a sum-product form is an undirected graph with
vertices corresponding to factor functions and edges corresponding to the variables.
The NFG of the posterior value (\ref{finerrep}), denoted by ${\cal G}^*$, is defined as follows.
For each factor of (\ref{finerrep}) such as 
\[
OR(x_{i,k}|_{k \in \alpha(i) } ),\  EQ(x_{k,j}|_{k \in \beta(j)} , u_j),\  \phi^{(\ell,b)}_j(u_j),
\]
a {\em factor node} is associated. In the following, we do not strictly distinguish a factor function 
from the corresponding factor node if there are no fear of confusion.
In a similar way, the variables in (\ref{finerrep})  such as $x_{i,j} \in \Gamma$  and $u_1,\ldots, u_n$  are assigned to 
edges. The rule for the edge connections is simple; if and only if a variable $x_{i,j}$ (resp. $u_i$) is an argument of
a factor function $f$, 
the edge $x_{i,j}$ (resp. $u_i$) is connected to the factor node $f$.
In other words, if and only if $f$ depends on $x_{i,j}$ (resp. $u_i$), the factor node $f$ connects to 
the edge $x_{i,j}$ (resp. $u_i$).
According to the semantics for NFGs introduced by Al-Bashabsheh and Mao \cite{Al-Bashabsheh},  all the edge variables
are assumed to be marginalized.

\begin{figure}[htbp]
\begin{center}
\includegraphics[scale=0.4]{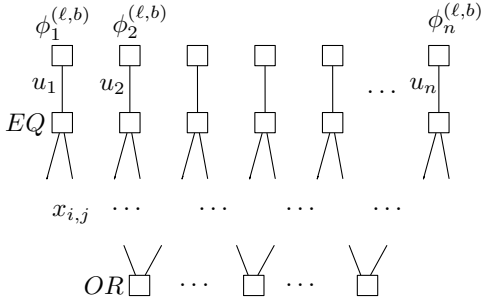}
\end{center}
\caption{Normal factor graph for the posterior value}
\label{nfg1}
\end{figure}

Figure \ref{nfg1} illustrates the NFG for the posterior value in (\ref{finerrep}).
We will refer the factor nodes corresponding to the OR (resp. EQ) function as {\em OR (resp. EQ) nodes}.

\subsection{Holographic transformation}
\label{holographic}
In our context, an NFG corresponds to  a posterior value in sum-product form.
A holographic transformation is a transformation of an NFG that preserves the marginal generating function.
In the following discussion, we will insert a pair of {\em dual factor nodes} into each edge connecting 
an OR node and an EQ node ({\em i.e., }  $x_{i,j}$). The pair of factor nodes is carefully designed not to
change the  posterior value.

\begin{figure}[htbp]
\begin{center}
\includegraphics[scale=0.3]{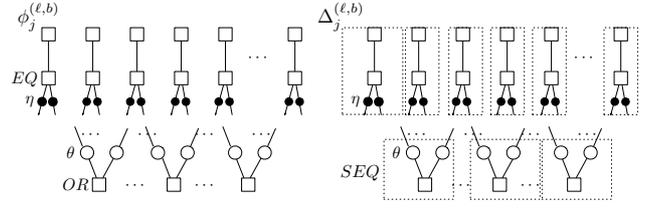}
\end{center}
\caption{Holographic transformation for NFG of posterior values}
\label{nfg3}
\end{figure}

Figure \ref{nfg3} is our blueprint that shows how we will proceed in this subsection.
In Fig.\ref{nfg3} (Left), for each edge $x_{i,j}$, 
a pair of dual nodes, $\theta$ and $\eta$, is inserted. 
These function nodes $\theta$ and $\eta$ are designed to satisfy the duality condition 
described later. Due to the duality condition on $\theta$ and $\eta$, 
the posterior value of this transformed NFG is the same as that of the original NFG. 
By grouping an EQ node and $\eta$ nodes connected to it, a new factor node  
$\Delta_j^{(\ell,b)}$ is created (Fig.\ref{nfg3} (Right)). 
In a similar way, combining an OR node and its incident $\theta$-nodes,
we obtain a new factor nodes that is called a {\em skewed EQ (SEQ) function}.
It should be emphasized that SEQ function has almost the same truth table as
that of EQ function. This fact is important to reduce the computational complexity 
to evaluate the posterior values.

In the following subsections, we will follow this blueprint and present 
details of dual factor nodes and new factor nodes.
These will be the basis of the proof of Theorem \ref{dualexpression}.

\subsection{Dual factor nodes}

Let us define $\theta: \{0,1\}^2 \rightarrow \{-1,0,+1\}$ by
\begin{equation}
\theta(i,j) \defeq 
\left\{
\begin{array}{ll}
0, & (i,j) = (0,0) \\
-1, & (i,j) = (0,1) \\
1, & (i,j) = (1,0) \\
1, & (i,j) = (1,1), \\
\end{array}
\right.
\end{equation}
and $\eta: \{0,1\}^2 \rightarrow \{-1,0,+1\}$ by
\begin{equation}
\eta(i,j) \defeq
\left\{
\begin{array}{ll}
1, & (i,j) = (0,0) \\
1, & (i,j) = (0,1) \\
-1, & (i,j) = (1,0) \\
0, & (i,j) = (1,1). \\
\end{array}
\right.
\end{equation}
It is trivial to check that 
these two functions $\theta$ and $\eta$ satisfies the duality condition 
\begin{equation}
\sum_{y \in \{0,1\} } \theta(x, y) \eta(y, w) = EQ(x,w).
\end{equation}
This condition guarantees that the posterior values of the NFG are unchanged 
if we inserted these function nodes into an edge corresponding to $x_{i,j}$ in Fig.\ref{nfg1} \cite{Al-Bashabsheh}.
This is because a pair of function nodes is equivalent to an EQ function which does not 
affect the consequence of the marginalization.

The next lemma tells that a sum-product form of an OR function and $\theta$ functions 
produces an SEQ function.
\begin{lemma}[SEQ function] \label{ORlemma}
For any $(y_1,\ldots, y_r) \in \{0,1\}^r$,  the following equality 
\begin{eqnarray} \nonumber
&&\hspace{-2.5cm}\sum_{x_1,x_2,\ldots, x_r}  OR(x_1,\ldots, x_r) \prod_{i=1}^r \theta(x_i, y_i)  \\ \label{thetaeq}
&= &
(-1)^{y_1(r+1)}EQ(y_1,\ldots, y_r)
\end{eqnarray}
holds.
\end{lemma}
Proof:
From the definition of $\theta$,  the right-hand side of (\ref{thetaeq}) can be evaluated as
\begin{eqnarray}\nonumber
&&\hspace{-1.5cm}\sum_{x_1,x_2,\ldots, x_r}  OR(x_1,\ldots, x_r) \prod_{i=1}^r \theta(x_i, y_i) \\  \nonumber
&=& \left[
((1,1)^{\otimes r} - (1,0)^{\otimes r}) 
\left(
\begin{array}{cc}
0 & -1 \\
1 & 1 \\
\end{array}
\right)^{\otimes r}
\right]_{y_1,\ldots, y_r} \\ \label{ttt}
&=& \left[
(1,0)^{\otimes r}- (0,-1)^{\otimes r}
\right]_{y_1,\ldots, y_r}. 
\end{eqnarray}
Note that $A^{\otimes r}$ represents the tensor power (Kronecker power) of a matrix $A$.
The row vector $(1,1)^{\otimes r} - (1,0)^{\otimes r}$ represents the truth table of 
$OR$ function as a row vector. The notation $[\mathbf{v}]_{a_1,\ldots, a_r}$ denotes
the $b$-th component of row (or column) vector $\mathbf{v}$ where $b=a_1+2^1 a_{2} + \cdots + 2^{r-1} a_r$.
If $r$ is odd, then the right-hand side of (\ref{ttt}) equals 
$
\left[
(1,0)^{\otimes r} + (0,1)^{\otimes r}
\right]_{y_1,\ldots, y_r}.
$
In this case, the claim of the lemma holds because 
$
\left[
(1,0)^{\otimes r} + (0,1)^{\otimes r}
\right]_{y_1,\ldots, y_r}
$
is the truth table of $EQ$ function.
If $r$ is even, 
the right-hand side of (\ref{ttt}) becomes
$
\left[
(1,0)^{\otimes r} - (0,1)^{\otimes r}
\right]_{y_1,\ldots, y_r},
$
which is equivalent to the right-hand side of (\ref{thetaeq}).
\hfill\qed

It can be seen that only simple tensor calculations are required to show the main claim of this lemma.
The right-hand side of (\ref{thetaeq}), $(-1)^{y_1(r+1)}EQ(y_1,\ldots, y_r)$, 
is referred to as the skewed EQ function that is denoted by $SEQ(y_1,\ldots, y_r)$.

The next lemma plays an crucial role for grouping factor nodes around an EQ node.
\begin{lemma}[Delta function] \label{EQlemma}
The function $\Delta_j^{(\ell, b)}(y_1,\ldots, y_r) $ can be expressed as
\begin{eqnarray} \nonumber
\Delta_j^{(\ell, b)}(y_1,\ldots, y_r)
&=& \sum_u \sum_{w_1,\ldots, w_r} EQ(u, w_1,\ldots, w_r)  \\
&\times&  \phi_j^{(\ell,b)}(u) \left(\prod_{k=1}^r \eta(y_k,w_k) \right)
\end{eqnarray}
for any $(y_1,\ldots, y_r) \in \{0,1\}^r, j \in [1,n], \ell \in [1,n], b \in \{0,1\}$.
\end{lemma}
Proof: The truth table of $EQ$ function of $(r+1)$-inputs is given by 
the column vector 
\[
\left(
\begin{array}{c}
1  \\
0 \\
\end{array}
\right)^{\otimes (r+1)}
+
\left(
\begin{array}{c}
0  \\
1 \\
\end{array}
\right)^{\otimes (r+1)}.
\]
From the definition of the function $\eta$, we have the following tensor expression:
\begin{eqnarray} \nonumber
&&\hspace{-1cm} \sum_{w_1,\ldots, w_r} EQ(u, w_1,\ldots, w_r)  \left(\prod_{k=1}^r \eta(y_k,w_k) \right) \\ \nonumber
&=& \hspace{-0.3cm}
\left[
\left(
\begin{array}{c}
1  \\
-1 \\
\end{array}
\right)
^{\otimes r}
\otimes
\left(
\begin{array}{c}
1  \\
0 \\
\end{array}
\right)
+
\left(
\begin{array}{c}
1  \\
0 \\
\end{array}
\right)^{\otimes r}
\otimes
\left(
\begin{array}{c}
0  \\
1 \\
\end{array}
\right)
\right]_{u,y_1,\ldots, y_r}.  
\end{eqnarray}

We here define $D_j^{(\ell,b)}(y_1,\ldots, y_r)$ as 
$
D_j^{(\ell,b)}(y_1,\ldots, y_r)
\defeq \sum_u \sum_{w_1,\ldots, w_r} EQ(u, w_1,\ldots, w_r)  \phi_j^{(\ell,b)}(u) \left(\prod_{k=1}^r \eta(y_k,w_k) \right).  
$
From the definition of $\phi_j^{(\ell, b)}(s)$, if $j \ne \ell$, we have
\begin{equation}
D_j^{(\ell, b)}(y_1,\ldots, y_r)
=
\left\{
\begin{array}{l}
1,\quad  y_1 = \cdots = y_r = 0 \\
(-1)^{\sum_{k=1}^r y_k} P_{X_j}(0),\quad  \mbox{otherwise}.
\end{array}
\right. 
\end{equation}
Otherwise ({\em i.e.,} $j = \ell$),  the equality 
\begin{equation}
D_j^{(\ell, b)}(y_1,\ldots, y_r) 
=
\left\{
\begin{array}{l}
P_{X_j}(b),\quad  y_1 = y_2 = \cdots = y_r = 0 \\
(-1)^{\sum_{k=1}^r y_k} P_{X_j}(0) \Bbb I[b=0], \mbox{otherwise}.
\end{array}
\right.
\end{equation}
is obtained. It is clear that $\Delta_j^{(\ell,b)} = D_j^{(\ell,b)}$ holds.
\hfill\qed

\subsection{Proof of Theorem \ref{dualexpression}}

We are now ready to prove Theorem \ref{dualexpression}. 
As described in Subsection \ref{holographic},
the NFG illustrated in Fig. \ref{nfg3} (Left) corresponds to the original posterior value 
due to the duality condition on $\theta$ and $\eta$.  By Lemmas \ref{ORlemma} and \ref{EQlemma},
the posterior value $a^{(b)}(\ell)$ can be rewritten as
\begin{eqnarray} \nonumber
a^{(b)}(\ell)  &=&  
\sum_{\Gamma'}  \left(\prod_{i=1}^m (-1)^{y_{i, k^*(i)} (\#\alpha(i)+1)}EQ(y_{i, k}|_{k \in \alpha(i)} )  \right) \\ \label{otherrep}
&\times& \left(\prod_{j =1}^n \Delta_j^{(\ell,b)} (y_{k,j}|_{k \in \beta(j)} ) \right),
\end{eqnarray}
where $\Gamma'$ is a set of variables defined by
\[
\Gamma' \defeq \{y_{i,j} \mid i \in [1,m], j \in [1,n], (z_i^{(2)},z_j^{(1)}) \in E^* \}
\]
and 
$
k^*(i) \defeq \min \{k \mid k \in \alpha(i)\}.
$
Note that the expression (\ref{otherrep}) follows the new factor node grouping described in Fig. \ref{nfg3} (Right).
In the non-vanishing summand of (\ref{otherrep}),  EQ function enforces that 
$y_{i, k}|_{k \in \alpha(i)}$ takes the same value for any $i$. In other words, 
all the edges emitted from a skewed EQ factor node should take the same value if the product 
in (\ref{otherrep}) is not zero.
This observation leads to the claim of Theorem \ref{dualexpression}.
\hfill\qed

\subsection{Numerical example}
The next example presents how this dual expression works. 

\begin{figure}[htbp]
\begin{center}
\includegraphics[scale=0.4]{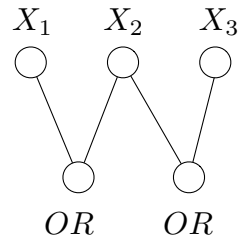}
\end{center}
\caption{Example of pooling graph}
\label{g1fig}
\end{figure}

Assume that $n=3, m=2$ and the pooling graph defined by $\alpha(1) = \{1,2\}, \alpha(2) = \{2,3\}$ (See. Fig. \ref{g1fig}).
The prior probabilities of objects are given as $P_{X_j}(0)=0.9, P_{X_j}(1)=0.1 (j \in [1,3])$.
Table \ref{g1} shows the prior probabilities $P_X(x_1,x_2,x_3) = \prod_{i=1}^3 P_{X_i}(x_i)$ and 
the indicator values $\Bbb I[OR(x_1,x_2)=1] \Bbb I[OR(x_3,x_3)=1]$.
We here focus on the case where $\ell = 2$.
From Table \ref{g1}, 
It is straightforward to evaluate the posterior values as 
\begin{eqnarray} \nonumber
a^{(0)}(2) &=& 0.009, \\ \nonumber
a^{(1)}(2) &=& 0.081 + 0.009 + 0.009 + 0.001 = 0.1.
\end{eqnarray}
\begin{table}
\begin{center}
\caption{Prior probabilities and indicator value}
\label{g1}
\begin{tabular}{ccc|cc}
\hline
$x_1$ & $x_2$ & $x_3$ & $P_{X}$ & Indicator value\\
\hline
0 & 0 & 0 & 0.729 & 0 \\
0 & 0 & 1 & 0.081 & 0 \\
0 & 1 & 0 & 0.081 & 1 \\
0 & 1 & 1 & 0.009 & 1\\
1 & 0 & 0 & 0.081 & 0 \\
1 & 0 & 1 & 0.009 & 1 \\
1 & 1 & 0 & 0.009 & 1 \\
1 & 1 & 1 & 0.001 & 1 \\
\hline
\end{tabular}
\end{center}
\end{table}
From the definition of $\Delta_j^{(\ell,b)}$, we obtain the following Delta functions:
\begin{eqnarray} \nonumber
\Delta_1^{(2, b)}(y) &=& \Delta_3^{(2, b)}(y) = 
\left\{
\begin{array}{ll}
1, & y = 0 \\
-0.9, & y= 1 
\end{array}
\right. \\ \nonumber
\Delta_2^{(2,0)}(y_1,y_2) &=&
\left\{
\begin{array}{ll}
0.9, & y_1=y_2=0 \\
0.9 \times (-1)^{y_1+y_2}   & otherwise
\end{array}
\right. \\ \nonumber
\Delta_2^{(2,1)}(y_1,y_2) &=&
\left\{
\begin{array}{ll}
0.1, & y_1=y_2=0 \\
0, & otherwise.
\end{array}
\right.
\end{eqnarray}
Table \ref{proddelta} presents the values of Delta function product for given $y_1$ and $y_2$,
and the value of $\sigma(y_1, y_2)$ defined by $\sigma(y_1, y_2) = (-1)^{y_1+y_2}$.
Due to Theorem \ref{dualexpression}, the posterior values $a^{(0)}(2)$ and $a^{(1)}(2)$ can be obtained as
\begin{eqnarray}  \nonumber
a^{(0)}(2) &=& \sum_{y_1,y_2} \sigma(y_1,y_2) \Delta_1^{(2,0)} \Delta_2^{(2,0)} \Delta_3^{(2,0)}  \\
&=& 0.9-0.81-0.81+0.729 = 0.009 \\ \nonumber
a^{(1)}(2) &=& \sum_{y_1,y_2} \sigma(y_1,y_2) \Delta_1^{(2,1)} \Delta_2^{(2,1)} \Delta_3^{(2,1)}  \\
&=& 0.1.
\end{eqnarray}
These values are exactly same as the values obtained from Table \ref{g1}.

\begin{table}[htbp]
\begin{center}
\caption{Values of the Delta function product}
\label{proddelta}
\begin{tabular}{cc|r|r|r}
\hline
$y_1$ & $y_2$ & $\Delta_1^{(2,0)} \Delta_2^{(2,0)} \Delta_3^{(2,0)}$ 
& $\Delta_1^{(2,1)} \Delta_2^{(2,1)} \Delta_3^{(2,1)}$ & $\sigma$ \\
\hline
0 & 0 & $1\times 0.9 \times 1$ & $1 \times 0.1 \times 1$ & +1 \\
0 & 1 & $1 \times (-0.9) \times (-0.9) $ & $1 \times 0 \times (-0.9)$ & -1 \\
1 & 0 & $(-0.9) \times (-0.9) \times 1 $ & $(-0.9)\times 0 \times 1$ & -1 \\
1 & 1 & $(-0.9) \times 0.9 \times (-0.9)$ & $(-0.9)\times 0 \times (-0.9)$ & +1 \\
\hline
\end{tabular}
\end{center}
\end{table}

\section{Bitwise MAP Estimation Algorithms}

From engineering point of view, the primal advantage of Theorem \ref{dualexpression}  is that 
it provides an efficient bitwise MAP estimation algorithm. 
If  the degrees of a node in a pooling graph is bounded by a constant,
exhaustive evaluation of posterior values based on Theorem \ref{dualexpression} requires 
$O(n^2 2^m)$-time (a simple implementation trick can reduce this time complexity down to $O(n 2^m)$).
If $m < n$, this bitwise MAP algorithm achieves exponential speedup compared with 
a naive bitwise MAP algorithm based on  (\ref{posterior}) with time complexity $O(n 2^n)$.
In a typical use of a non-adaptive group testing,
the number of tests is much smaller than the number of objects; {\em i.e.}, $M << N$.
This implies that a situation satisfying $m < n$ is fairly common. 

\section*{Acknowdegement}
The authors would like to express their sincere appreciation to Ryuhei Mori 
for directing our interest towards holographic transformation.
This work was  supported by JSPS Grant-in-Aid for Scientific Research (B) Grant Number 25289114.

\end{document}